\documentclass[twocolumn]{aa}
\usepackage{graphicx}
\usepackage{amssymb}
\usepackage{amsmath}
\usepackage{float}
\usepackage{txfonts}
\usepackage{gensymb}
\usepackage{hyperref}
\usepackage{subfig}
\usepackage{threeparttable}
\usepackage[normalem]{ulem}
\usepackage[figuresright]{rotating}
\usepackage{longtable}
\usepackage{titlesec}

\begin{document}
	
\authorrunning{Hu, Q., et al.}

\titlerunning{The morphological stability of open clusters}
	
\title{3D projection analysis: characterizing the morphological stability of nearby open clusters}
	
\author{Qingshun Hu \inst{1}, Songmei Qin \inst{2,3,4}, Yangping Luo \inst{1} and Yuting Li \inst{1}}
	
\institute{School of Physics and Astronomy, China West Normal University, No. 1 Shida Road, Nanchong 637002, People's Republic of China, (\email{qingshun0801@163.com}) \label{inst1} \and Key Laboratory for Research in Galaxies and Cosmology, Shanghai Astronomical Observatory, Chinese Academy of Sciences, 80 Nandan Road, Shanghai 200030, China \label{inst2} \and School of Astronomy and Space Science, University of Chinese Academy of Sciences, No. 19A, Yuquan Road, Beijing 100049, China \label{inst3} \and Institut de Ci\`encies del Cosmos, Universitat de Barcelona (ICCUB), Mart\'i i Franqu\`es 1, 08028 Barcelona, Spain \label{inst4} }
	
\date{Received 25 June 2024 / Accepted 06 December 2024}

\abstract{The study of open cluster morphology is pivotal for exploring their formation and evolutionary processes.}
{We manage to assess the morphological stability of 105 nearby open clusters within tidal radii on the X-Y, X-Z, and Y-Z planes of the heliocentric Cartesian coordinate system, utilizing member catalogs from the literature. Meanwhile, we also delve into factors potentially impacting the clusters' morphological stability on these projection planes.}
{We used the rose diagram method by constructing the 3D projected morphology of sample clusters to quantify the morphological stability of their 3D projected morphology.}
{Our analysis indicates there is a demonstrated linear positive correlation between the number of sample clusters' member stars within tidal radii and their morphological stability in different 3D projection planes. This may suggest that the more member stars there are within the tidal radius of a cluster, the stronger its own gravitational binding capacity is, resulting in strong morphological stability. We find a direct link between the clusters' morphological stabilities in the X-Z plane within tidal radii and their velocity dispersion in the same plane, suggesting that the morphological stabilities in the X-Z plane may be dependent on internal dynamics. Moreover, the morphological stability of the clusters' 3D projection is influenced by their spatial positions along the Y axis, but not linearly, indicating that the environmental changes at the clusters' location may affect their morphological stability. Likewise, specific external forces can have an effect on their morphological stability.}
{This research introduces a novel perspective for understanding the morphological stability of open clusters, with a particular focus on their 3D projected morphologies.}


\maketitle{}


\section{Introduction}

Open clusters are typically born concurrently in the same molecular cloud \citep{lada03}. Initially, they are at the embedded stage \citep{pelu12}, a period characterized by intense gas dissipation. Post this stage, only a fraction of these clusters survive, with many disintegrations during the embedded phase \citep{lada03, good06}. The survivors often retain the morphology as their early stage \citep{krau20}. Over time, the clusters' star populations and spatial structures evolve due to some physical processes, including gravitational interactions among members, stellar evaporation \citep{port10, rein23}, and external perturbations \citep{spit58, port10} such as Galactic tidal forces \citep{hegg03, erns15, mcdi22, ange23}, gravitational perturbations from giant molecular clouds, and encounters with these molecular clouds. These evolutionary changes are reflected in the clusters' morphology, offering critical insights into their formation and evolution \citep{chen04, zhai17, hu21b}.

Decades of research on the morphology of open clusters, such as the works of \citet{cart04}, \citet{chen04}, \citet{sant05}, \citet{khar09}, \citet{zhai17}, \citet{dib18}, \citet{hete19}, \citet{hu21a}, \citet{hu21b}, \citet{tarr22}, and \citet{zhon22}, have culminated in a coherent understanding of their archetypal structure, notably the core-corona structure \citep[e.g.][]{arty66, khol69, chen04, mein21} or other morphologies like tidal tails \citep{mein19a, hunt23}; strings \citep{koun19}; rings \citep{cant19}; streams \citep{mein19b}; snakes \citep{tian20, wang22}; pearls \citep{coro22}. Capitalizing on the high precision data from the Gaia mission, \cite{pang21} examined the 3D ellipsoidal structures within the tidal radii of certain open clusters. More recently, \cite{hu23} identifies the layered structure of nearby open clusters in 3D space, using the rose diagram overlay technique.

It is well known that the morphology of open clusters does not remain constant, but changes continuously over time. The parameter measuring the likelihood of morphological changes in open clusters can be defined as morphological stability that can be expressed in various ways, such as the morphological coherence proposed by \citet{hu24}, and the core instability index defined by \citet{hu23}. The greater the morphological stability of a cluster, the greater its ability to maintain its own morphology unaltered by the external environment, and vice versa. Besides, the morphological stability may, to some extent, affect how long open clusters continue to live because clusters that are prone to change their morphology are theoretically more likely to collapse \citep[see the \texttt{Blanco~1} in their Fig.~6]{hu21b}. Although the more clear perception of open cluster morphology from two dimension (2D) to 3D \citep{zhai17,hu21a,hu21b,tarr22,zhon22,hu23,hu24}, its morphological stability and the factors influencing the morphological stability remain elusive. Thus, the profound exploration of open cluster morphology is restricted. Besides, unraveling the determinants of morphological stability is imperative for advancing our comprehension of these stellar population and their evolution.

Our recent research \citep{hu24} indicates that the 3D morphological stability of open clusters does not appear to be substantially influenced by the external environment. Nonetheless, it is plausible to hypothesize that the 3D morphology of clusters could be impacted by external factors, potentially due to a counterbalancing of various external influences on their 3D structures. And, of course, it is possible that this is because part of their sample clusters contains tidal structures and others do not. In addition, the tidal structure of the clusters is outside of the gravitational boundaries of the clusters, it can be regarded as part of the external environment, except that it borders the gravitationally bounded boundaries of the clusters. Therefore, in this study, we introduce the use of rose diagrams developed by \cite{hu23} to dissect these complex superimposed effects and focus only on the open clusters within their tidal radii. Specifically, we employed rose diagrams to characterize the stability of the projected morphology of open clusters within the 3D space. Additionally, we investigate the correlation between the stabilities of these projected morphologies and the external environment.

The structure of this paper is as follows: Sect.~2 outlines the data sources and Sect.~3 details the methodology employed in our work. Sect.~4 presents the findings of our study. To conclude, Sect.~5 provides a summary of our results and presents the conclusion.

\section{Data}

\label{data}

Our sample clusters are selected from the comprehensive catalog of 2492 open clusters compiled by \citet{van23}. \citet{cant20} and \citet{cast22} provide a list of open cluster member stars by an unsupervised classification scheme UPMASK method \citep{kron14, cant18} and a density based clustering algorithm DBSCAN method \citep{este96}, respectively. Building upon their cluster catalog, \citet{van23} utilized Gaia Data Release~3 (Gaia-DR3) \citep{gaia23} to recover members of the 2492 open clusters. This cluster catalog is elaborated using a machine learning based tool designed to discern the distribution of reliable members within known open clusters, thereby enhancing the recovery of their member stars \citep{van23}. This effort includes the identification of faint members (with G$>$18~mag) that are not previously detected by \citet{cant20} and \citet{cast22}. In this work, we filter out 105 clusters from 2492 open clusters by applying the criterion of a cluster center parallax $\varpi$ greater than or equal to~2~mas. These 105 open clusters are our initial sample clusters. Furthermore, the sample clusters exhibit a wide age range, spanning from a few million years to several gigayears. 

In the current study, we then meticulously selected the sample clusters' member stars with a membership probability of 0.5 or higher to ensure the reliability of member stars for our sample clusters. Therefore, our final sample contains only the clusters with member stars with membership probability~$\geq$~0.5 in the initial sample.

\section{Method}

A portion of the sample clusters have tidal structures and these are not within the tidal radius of the sample clusters while being released from the gravitational binding of the clusters. This study focuses on the morphology of the clusters within the tidal radius. Therefore, we exclude the members outside the tidal radius of the sample clusters. To achieve this, we applied the \texttt{King} model \citep{king62} to fit the projected distribution of the sample clusters within a 2D celestial coordinate system to derive their tidal radii. This process was facilitated by the Python code available in the \texttt{gaia$\_$oc$\_$amd} repository on GitHub\footnote{\url{https://github.com/MGJvanGroeningen/gaia_oc_amd}}, as provided by \citet{van23}. We used this code to fit and obtain the tidal radii of sample clusters in the present work. Final, we singled out all the member stars of each cluster that lie within the tidal radius. The equation of the \texttt{king model} used in the code is as follows:

\begin{equation}
	\centering
	f(r) = k \cdot (\frac{1}{\sqrt{1+(r/r_{c} )^{2} } } - \frac{1}{\sqrt{1+(r_{t}/r_{c} )^{2}}} )^{2} + c,
\end{equation}

where $k$, $r_{c}$, $r_{t}$, and $c$ are a quantity that relates to the central density of clusters, a core radius, a tidal radius, and a background density, respectively. The meaning of these four parameters can be also seen in the function (\texttt{king\_model}) of the subpackage (\texttt{diagnostics}) of the \texttt{gaia$\_$oc$\_$amd}.

\begin{figure}[ht]
	\centering
	\includegraphics[width=82mm]{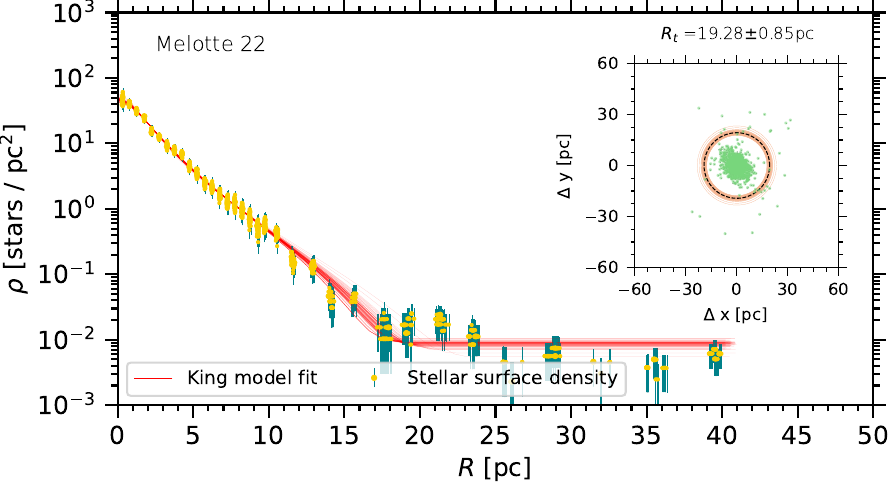}
	\includegraphics[width=82mm]{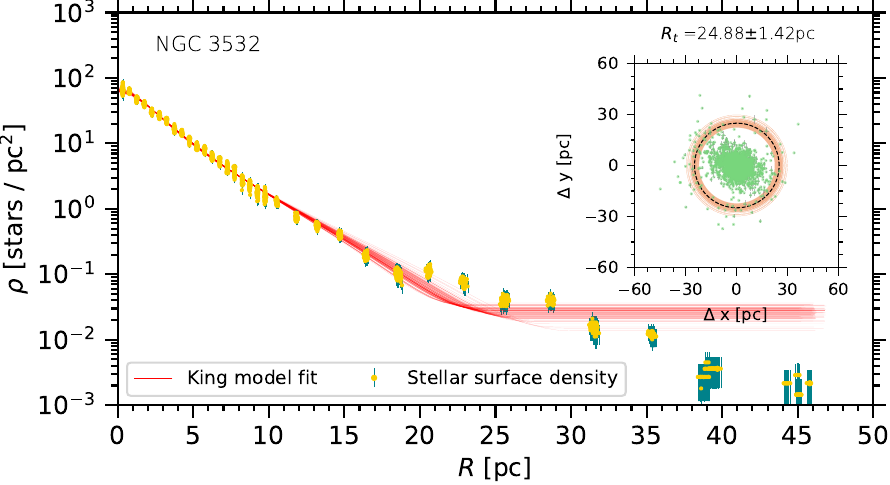}
	\caption{Surface density profiles of the two sample clusters (Top: Melotte~22 and Bottom: NGC~3532) and their projection distributions in the 2D celestial coordinate frame. The radii of the colored circles in the subplots are the tidal radii of the sample clusters, derived from 100 times of \texttt{King} model fitting. The black dashed circle in each subplot is plotted based on the mean of the tidal radii. The green dots with gray error bars represent the sample cluster member stars, with the gray error bars being the errors of the coordinates of these members. The red solid lines denote the surface density profiles of the two clusters.}
	\label{Density_profile}
\end{figure}

In addition, we used a Monte Carlo method to calculate the error in the tidal radius of each sample cluster. We first repeated the sampling 100 times for each member star of each sample cluster by taking the measured values of the coordinate positions (\texttt{RA} and \texttt{DEC}) of the member stars of each sample cluster as the mean value of the Gaussian distribution sampling, and their measurement errors as the variance. The \texttt{king model} was then fitted 100 times to each sample cluster to obtain 100 fitted values for the tidal radius. Finally, the mean of the 100 tidal radius values for each cluster was taken as the tidal radius, and its standard deviation was taken as the error in the tidal radius. In addition, the medians of the relative errors of the coordinate positions of our sample clusters are about 0.06\% (\texttt{RA}), and 0.19\% (\texttt{DEC}), respectively. Similarly, the median of the relative errors of the tidal radii we fitted is about 18.91\%, with the median of tidal radii errors being about 8.44~pc. Figure~\ref{Density_profile} presents the \texttt{King} model fits for two representative clusters, Melotte~22 and NGC~3532, along with their respective projection distributions in the 2D celestial coordinate system. From these fits, we determined the tidal radii (denoted by the symbol ``$R_{t}$'') to be 19.28~$\pm$~0.85~pc for Melotte~22 and 24.88~$\pm$~1.42~pc for NGC~3532.

Furthermore, we used the rose diagram method to construct the 3D morphological projections of our sample clusters. To quantitatively characterize the 3D projected morphology of our sample clusters, we first determined their 3D spatial distribution within the heliocentric Cartesian coordinate system (X, Y, Z) \footnote{The reference frame is an XYZ Cartesian coordinate system, with the Sun at its center. The positive X axis is oriented from the Sun's projected position on the Galactic midplane toward the Galactic center. The positive Y axis aligns with the direction of the Galactic rotation, while the positive Z axis extends toward the north pole of the Galaxy.}. The coordinates (X, Y, and Z) for each member of the sample clusters were computed using the Python Astropy package \citep{astr13, astr18}.

However, this direct computation can result in an apparent stretching of the clusters in 3D space along the line of sight direction, due to the inversion of the measured parallaxes. Given that Gaia's parallax measurements are subject to symmetric errors \citep[see, e.g.,][]{bail15, luri18, carr19}, the resulting distance errors manifest as an asymmetric distribution when the measured parallaxes are inverted. To address this, we employed a Bayesian distance correction model, a method increasingly adopted in the literature \citep{carr19, pang21, ye21, qinm23, hu23}, to correct the distance discrepancies in our study. The Bayesian model we adopted is the same as that described in Appendix A of \citet{hu24}. We infer readers to see more details about the model in \citet{hu24}.

\begin{figure}[ht]
	\centering
	\includegraphics[width=86mm]{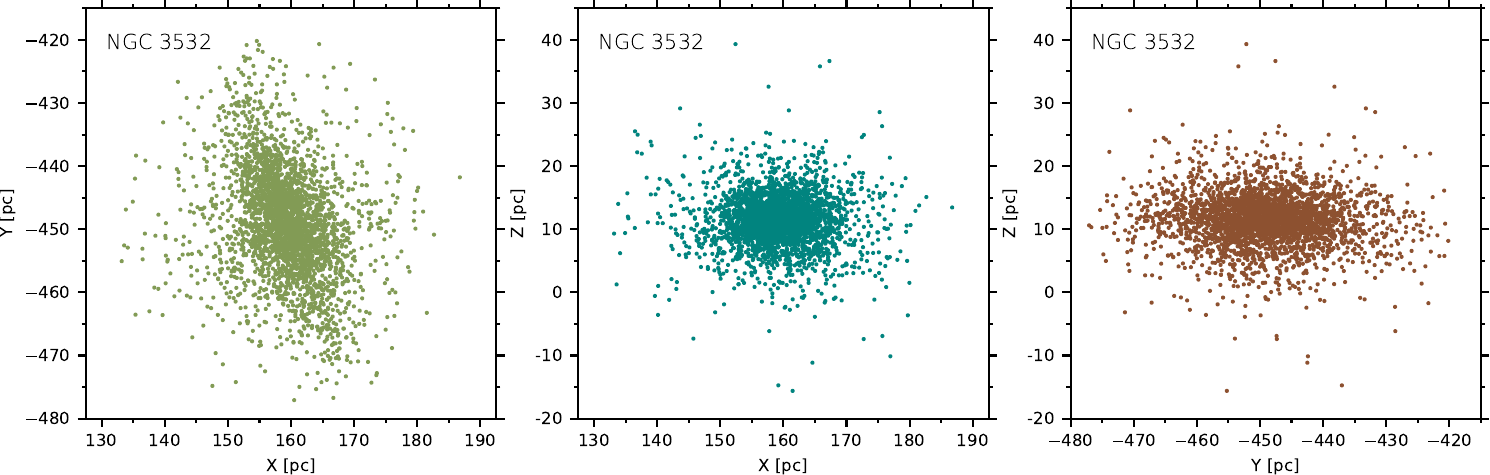}
	\caption{Projection distributions within tidal radii of the sample cluster (NGC~3532) on the X-Y plane (left), X-Z plane (middle), and Y-Z plane (right). Different colored dots in panels represent its members.}
	\label{3D_projection}
\end{figure}

After applying distance corrections to the member stars of all sample clusters, we derived their 3D projections within the heliocentric Cartesian coordinate system, including the X-Y, X-Z, and Y-Z planes. Figure~\ref{3D_projection} illustrates the 3D projection distribution of NGC~3532 as an example. Subsequently, we employed rose diagrams previously utilized by \cite{hu23}, to quantitatively characterize the 3D projected morphology of the sample cluster. This approach converts the dispersed, amorphous, and irregular distributions of sample member stars into normalized, quantifiable distributions.

The procedure for constructing the rose diagram of the 3D projection distribution for the sample clusters follows the methodology outlined by \citet{hu23}. Using the median distance from each star to the cluster's center and star counting techniques, we developed rose diagrams for all samples across the three distinct projection planes. The general steps are as follows. 1) Each 3D projection distribution of the sample cluster members is divided into 12 sectors, each spanning an angle of 30 degrees, following the guidelines set by \citet{hu23}, as depicted in Fig.~\ref{Rose_projection}. 2) The radius (r$_{i}$) of each sector is calculated based on two parameters: the number (n$_{i}$) of member stars within each sector and the median distance (d$_{i}$) of those stars from the sector's center (cluster's center). 3) The covariates (n$_{i}$ and d$_{i}$) for each sample cluster are normalized by the following equation:

\begin{figure}[ht]
	\centering
	\includegraphics[width=86mm]{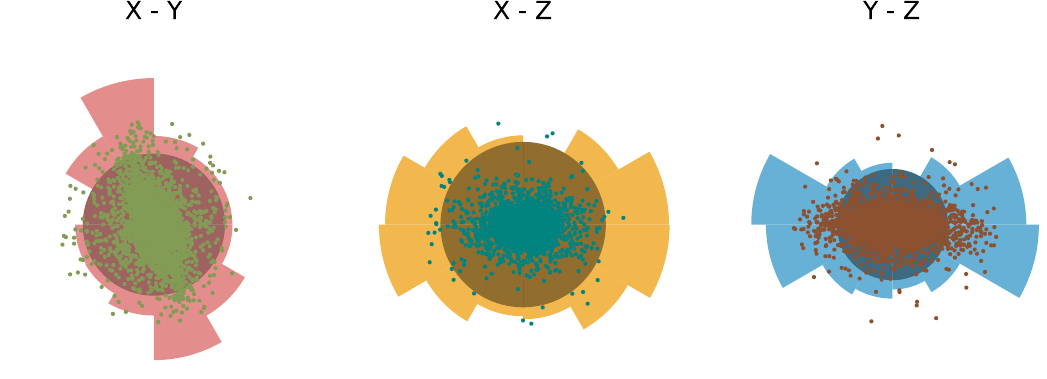}
	\caption{Rose diagrams within tidal radii of the cluster (NGC~3532) on the X-Y plane (left), X-Z plane (middle), and Y-Z plane (right) and the projected distributions of its members within tidal radii on these planes. The dark shaded circular area in each projection plane corresponds to its central region area ($S_{core_{XY}}$, or $S_{core_{XZ}}$, or $S_{core_{YZ}}$). The different colored dots are the same as those in Fig.~\ref{3D_projection}.}
	\label{Rose_projection}
\end{figure}

\begin{equation}
	\centering
	n^{norm}_{i} = \frac{n_{i}}{max(n_{i})},  \qquad
	d^{norm}_{i} = \frac{d_{i}}{max(d_{i})}. 
\end{equation}

Subsequently, we established the relationship between the normalized covariates (n$^{norm}_{i}$ and d$^{norm}_{i}$) and the radius (r$_{i}$) of each sector, as detailed in Eq.~\ref{radius_equation},

\begin{equation}\label{radius_equation}
	r_{i} = \sqrt{(n^{norm}_{i})^2 + (d^{norm}_{i})^2}.
\end{equation}

In this context, $r_{i}$ is dimensionless quantity that includes $r^{XY}_{i}$, $r^{XZ}_{i}$, and $r^{YZ}_{i}$, with the index $i$ ranging from 1 to 12. This equation is the same as the Eq.~2 of \citet{hu23}

Based on these expressions, we are able to deduce the parameters ($r^{XY}_{i}$, $r^{XZ}_{i}$, and $r^{YZ}_{i}$) for each sample's projection distribution. Then we constructed the rose diagrams for the 3D projection distributions of each sample cluster.

It can be imagined that the central region of a cluster's 3D projected shape will usually be concentrated into an approximately circular system due to its internal gravitational binding, while the periphery of the region exhibits an irregular distribution due to the cluster's perturbation by the external environment. According to this principle, we can quantitatively assess the stability of the projected morphology of clusters on the basis of the rose diagrams of their 3D projected morphology. To achieve this, we referred to the instability index ($\eta$) for 3D layered structures as defined in \citet{hu23}, specifically, their Eqs.~3-5. In the present study, we calculate the morphological stability for each projected distribution individually, employing the following equations, analogous to those used by \citet{hu23},

\begin{equation}
	S_{core} = \pi \cdot min(r_{i})^2,
\end{equation}

\begin{equation}
	S_{outer} = \sum_{i=1}^{i=12} (\frac{\pi \cdot r_{i}^{2}}{12}) - S_{core},
\end{equation}

where $S_{core}$ is comprised of $S_{core_{XY}}$, $S_{core_{XZ}}$, and $S_{core_{YZ}}$, similarly, with $S_{outer}$ containing $S_{outer_{XY}}$, $S_{outer_{XZ}}$, and $S_{outer_{YZ}}$. These quantities are dimensionless and serve to describe the morphology of the sample clusters. The $S_{core}$ represents the area of the central region of the 3D projected distribution within the tidal radii of the sample clusters, while the $S_{outer}$ refers to the area of the peripheral regions within the tidal radii beyond these cores. Utilizing the aforementioned equations, we obtained the morphological stabilities for the 3D projected distributions of all sample clusters. 

Similar to the error calculation for the tidal radii, we also used the Monte Carlo method to estimate the errors of the morphological stabilities. We first took the tidal radius obtained as the mean of the Gaussian sampling, and its error as the variance. A tidal radius is obtained by sampling each sample cluster once, and then the member stars of this cluster within that tidal radius are filtered out, and their coordinates (X, Y, Z) are substituted into our rose diagram code to calculate the area of the center region as well as the peripheral area of this cluster on each projection plane, and finally, the morphological stability of this sample cluster can be computed for all the three projected planes. Repeated sampling of the tidal radius of each sample cluster in turn, and finally the mean value of the morphological stabilities of each sample cluster is taken as the morphological stability of each sample cluster, and its standard deviation is taken as the error of the morphological stability.

In this work, we consider the sample clusters with $S_{core}$ = 0 as highly unstable clusters in the projected morphology. Because they have not a regular core, but almost an irregular shape within their tidal radii. This may indicate that they are easily tidally disrupted in the future. There are two sample clusters with $S_{core}$ = 0 in our work, which indicates that there is at least one sector without member distribution in their 3D projected planes. The two clusters are \texttt{Alessi~13} and \texttt{UPK~88}, with 33 and 49 member stars within tidal radii, respectively, where the first cluster has $S_{core}$ = 0 in all projective planes, with the second cluster having $S_{core}$ = 0 in the X-Y and X-Z projective planes only. It is noteworthy that clusters exhibiting  $S_{core}$ = 0 in their 3D projected distribution are excluded from the subsequent analysis, because such clusters, characterized by a sparse and uneven distribution of member stars, may skew the results. By constructing rose diagrams for each projection distribution of the sample clusters in 3D space, as depicted in Fig.~\ref{Rose_projection} (only one time of sampling), we can observe that NGC~3532, for instance, has distinct rose diagrams for the X-Y, X-Z, and Y-Z projection planes. These diagrams delineate the central core areas (represented by $S_{core}$, shown in a dark gray) and the irregular outer regions (represented by $S_{outer}$, indicated by the area outside the dark gray field) for each projection plane.

\begin{figure}
	\centering
	\includegraphics[angle=0,width=86mm]{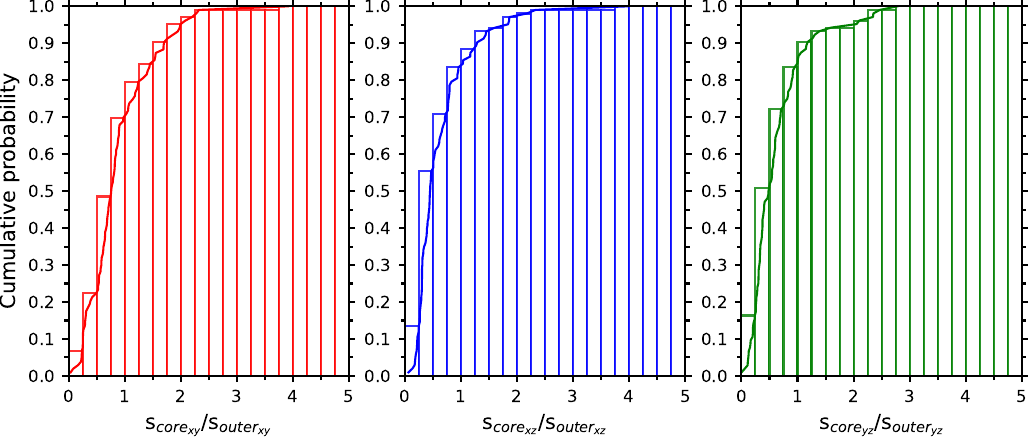}
	\caption{Histogram of cumulative distribution (left: X-Y plane, middle: X-Z plane, and right: Y-Z plane) of morphological stability of 3D morphological projections of sample clusters with their cumulative distribution profile lines (different colored curves).}
	\label{fig: ratio_CDF}
\end{figure}

\citet{hu23} introduces the concept of the instability index ($\eta$) for the 3D layered structure of open clusters, revealing a significant negative correlation between this index and the number of member stars within clusters. Furthermore, \cite{hu24} first defines the morphological stability of the 3D ellipsoidal structure of open clusters, characterized by morphological dislocations \citep{hu21b}. They discover a correlation between morphological stability and cluster star count, yet find no indication that external environmental factors influence this stability. We hypothesize that while the external environment likely impacts cluster morphological stability, these effects may counterbalance each other when not considering sample effects.

To investigate this hypothesis, we redefine the morphological stability as $S_{core}$/$S_{outer}$ in this study, which is essentially the inverse of the instability index as presented by \citet{hu23}, as seen in their Eq.~5. Because the previous core instability index measures the 3D morphological stability of open clusters, i.e., the larger the core instability index, the worse the 3D morphological stability of the cluster. In this work, we study the morphological stability of open clusters in the 2D projection plane, i.e., it is the 2D morphological stability. The essence of morphological stability is the same in both 2D and 3D, but the expression form is different. The former is the area common to the superposition of the three projection planes minus the area of its smallest core circle divided by the area of the smallest core circle, and the latter is the area on a single projection plane minus the area of its smallest core circle divided by the area of this smallest core circle. This redefinition is based on the projection morphologies of the 3D structure of sample clusters, such as the projected distributions illustrated in Fig.~\ref{3D_projection}.

In this way, we can determine the central core and irregular outer area sizes for all sample clusters on three distinct projection planes. Furthermore, we can calculate the morphological stabilities for the 3D projection distributions of the sample clusters. Figure~\ref{fig: ratio_CDF} shows the cumulative histogram of the morphological stability of 3D morphological projections of sample clusters on the X-Y, X-Z, and Y-Z planes. We assumed that the larger the regular internal area of a sample cluster, the stronger its self-gravitational binding capacity. This is because typically open clusters have a more aggregated distribution of inner member stars than outer member stars. If the regular internal area of a sample cluster is larger than its irregular external area, i.e., its morphological stability is larger than 1, then we can assume that the sample cluster has a relatively stable morphological structure, and vice versa.
	
We here can regard the morphological stability equal to 1 as a threshold value which indicates the regular internal area of a sample cluster being equal to its irregular external morphological area. We can see from Fig.~\ref{fig: ratio_CDF} that the morphological stability of most of the sample clusters (70.87\% for X-Y plane, 84.47\% for X-Z plane, and 84.62\% for Y-Z plane) is less than 1 no matter which projection plane they are on. This means that nearly 3/4 of our sample clusters on three projected planes have a relatively unstable morphology. Because their regular core area is smaller than their irregular outer area. Moreover, we find the sample clusters with morphological stability greater than 1 almost reached a median age (\texttt{Logt} = 7.93~$\pm$~0.52 for them on the X-Y plane, \texttt{Logt} = 7.86~$\pm$~0.57 on the X-Z plane, and \texttt{Logt} = 7.89~$\pm$~0.61 on the Y-Z plane). \citet{pisk18} find a lack of star clusters with age greater than 1~Gyr (\texttt{Logt} = 9.0). This phenomenon is also present in our sample. The older sample clusters must have evolved from middle-aged clusters. However, not all middle aged sample clusters survive to the old age stage. Sample clusters that survive to old age should be those with relatively stable morphology, while sample clusters with relatively unstable morphology may be disintegrated before they reach old age. Therefore, we can predict that these sample clusters with morphological stability greater than 1 have a potential ability to continue to survive into old age such as \texttt{Logt} = 9.0. The fundamental parameters of the sample clusters, their corresponding morphological stability parameters, and the errors of these parameters are collated in Tables~\ref{table:parameters_one},~\ref{table:parameters_two}, and~\ref{table:parameters_three} in the Appendix.

\section{Results}

\subsection{Morphological stability of the 3D projected distribution of sample clusters vs. the number of members and ages}

\begin{figure*}
	\centering
	\includegraphics[angle=0,width=50mm]{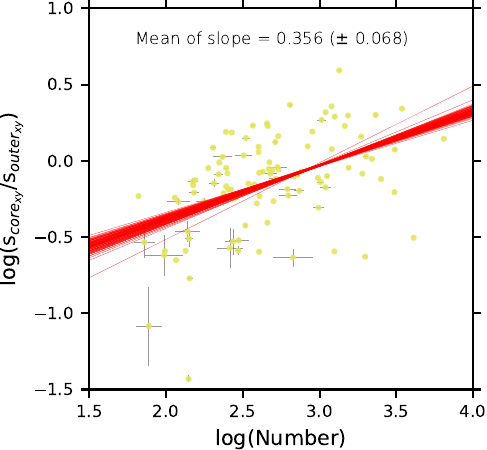}
	\includegraphics[angle=0,width=50mm]{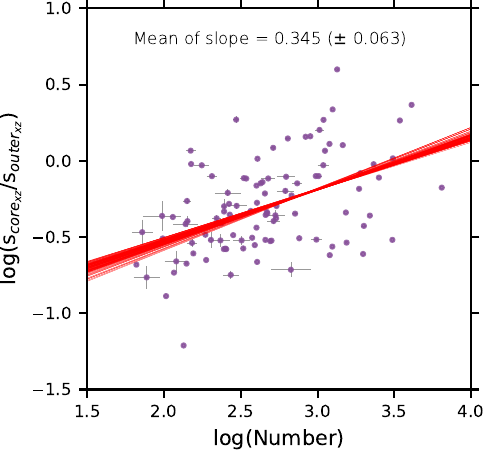}
	\includegraphics[angle=0,width=50mm]{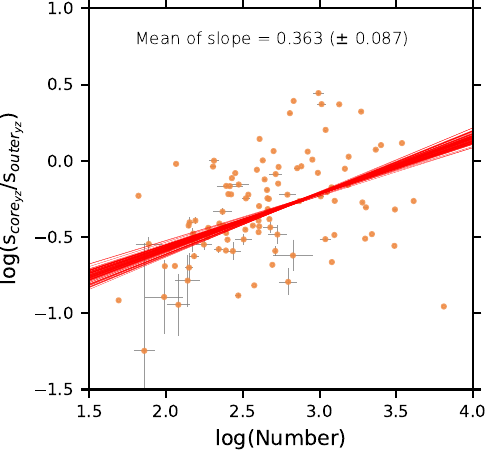}
	\caption{Logarithm of the morphological stability of sample clusters on three different projections within tidal radii vs. logarithm of the number of the sample cluster members within tidal radii. The different colored scatters with gray error bars (left: X-Y plane, middle: X-Z plane, and right: Y-Z plane) indicate the sample clusters, with the horizontal and vertical gray error bars representing the errors on the horizontal and vertical axes, respectively. The solid red lines are linear fit lines, which were obtained by sampling the parameters distributed in the plot 100 times within their error ranges using a Gaussian distribution, and linearly fitting the parameter distributions for each of these 100 times of samples. The slope in each label is the mean of the slopes of the 100 linear fits, with the value in the parenthesis being the mean of the standard deviation of the slopes of the 100 linear fits. It should be noted that the number of samples in each panel is not equal to that of all samples. Because we removed those samples with the central core area ($S_{core_{xy}}$, or $S_{core_{xz}}$, or $S_{core_{yz}}$) being zero.}
	\label{fig: ration_N}
\end{figure*}

\begin{figure*}
	\centering
	\includegraphics[angle=0,width=50mm]{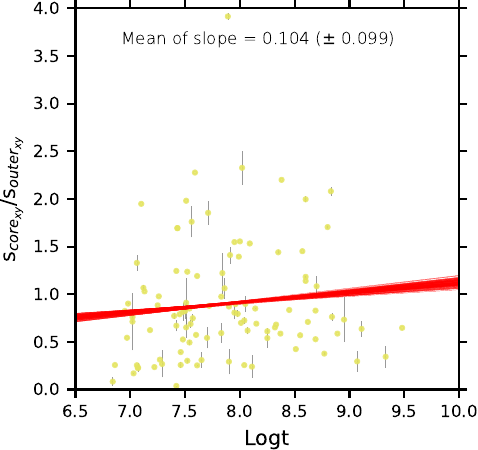}
	\includegraphics[angle=0,width=50mm]{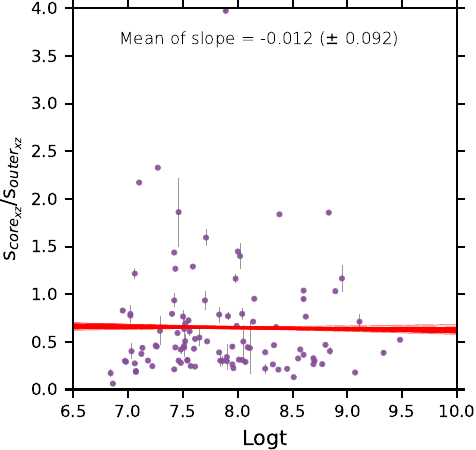}
	\includegraphics[angle=0,width=50mm]{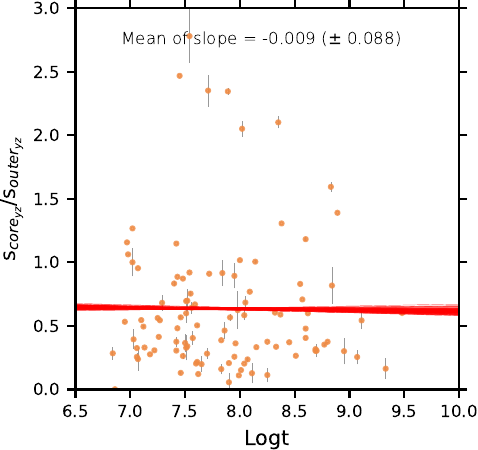}
	\caption{Morphological stability of sample clusters on three different projections within tidal radii vs. logarithm of the age of the sample clusters. The meaning of the values in the labels, symbols, and curves of the picture are the same as those of Fig.~\ref{fig: ration_N}. The age parameters of the sample clusters are from \citet{van23}. The solid red lines are linear fit lines. It should be noted that the number of samples in each panel is not equal to that of all samples. Because we removed those samples with the central core area ($S_{core_{xy}}$, or $S_{core_{xz}}$, or $S_{core_{yz}}$) being zero.}
	\label{fig: ration_Age}
\end{figure*}

\begin{table}[ht]\scriptsize
	\caption{Correlation information for sample clusters.}
	\centering
	
	\label{table:Correlation}
	\begin{tabular}{c c c}
		\hline\noalign{\smallskip}
		\hline\noalign{\smallskip}
		Variables & Pearson & $p$-value \\
		(1) & (2) & (3) \\
		\hline\noalign{\smallskip}
		log($S_{core_{xy}}$/$S_{outer_{xy}}$) vs. log(Number) & 0.46~$\pm$~0.01  & 1.45~$\times$~$10^{-6}$~$\pm$~1.80~$\times$~$10^{-6}$ \\
		log($S_{core_{xz}}$/$S_{outer_{xz}}$) vs. log(Number)  & 0.48~$\pm$~0.02 & 5.51~$\times$~$10^{-7}$~$\pm$~5.94~$\times$~$10^{-7}$  \\
		log($S_{core_{yz}}$/$S_{outer_{yz}}$) vs. log(Number) & 0.38~$\pm$~0.02  & 9.74~$\times$~$10^{-5}$~$\pm$~1.23~$\times$~$10^{-4}$ \\
		$S_{core_{xy}}$/$S_{outer_{xy}}$ vs. logt & 0.10~$\pm$~0.01  & 3.00~$\times$~$10^{-1}$~$\pm$~7.00~$\times$~$10^{-2}$ \\
		$S_{core_{xz}}$/$S_{outer_{xz}}$ vs. logt  & -0.01~$\pm$~0.01 & 8.90~$\times$~$10^{-1}$~$\pm$~7.00~$\times$~$10^{-2}$  \\
		$S_{core_{yz}}$/$S_{outer_{yz}}$ vs. logt& -0.01~$\pm$~0.01  & 9.10~$\times$~$10^{-1}$~$\pm$~6.00~$\times$~$10^{-2}$ \\
		$S_{core_{xy}}$/$S_{outer_{xy}}$ vs. $\sigma_{V_{xy}}$ & 0.04~$\pm$~0.02  & 6.58~$\times$~$10^{-1}$~$\pm$~1.11~$\times$~$10^{-1}$ \\
		$S_{core_{xz}}$/$S_{outer_{xz}}$ vs. $\sigma_{V_{xz}}$  & 0.30~$\pm$~0.01 & 2.19~$\times$~$10^{-3}$~$\pm$~9.82~$\times$~$10^{-4}$  \\
		$S_{core_{yz}}$/$S_{outer_{yz}}$ vs. $\sigma_{V_{yz}}$ & -0.12~$\pm$~0.01  & 2.35~$\times$~$10^{-1}$~$\pm$~5.58~$\times$~$10^{-2}$ \\
		\hline\noalign{\smallskip}
	\end{tabular}
	\tablefoot{Column~(1) represents the name of variables. Columns~(2) and ~(3) represent Pearson's correlation coefficients and probabilities for rejecting the null hypothesis that there is no correlation, respectively. In general, a correlation with a p-value of less than 0.05 indicates that the correlation is reliable.}
	\flushleft
\end{table}

In our previous study \citep{hu23}, we find a linear relationship between the morphological stability of open clusters and the number of their member stars. In this work, since we investigate the morphological stability of the 3D projected distribution within tidal radii of open clusters, we wonder if there is also a linear relationship between the morphological stability on different projections and the number of member stars of the cluster in this study. We constructed a distribution diagram correlating the morphological stability of these projections with the number of member stars, as depicted in Fig.~\ref{fig: ration_N}. We can see from all three panels of Fig.~\ref{fig: ration_N} that there are three of positive linear correlations between the morphological stability and the number of clusters' members. The correlation parameters can be seen in Table.~\ref{table:Correlation}. This correlation aligns with our previous findings \citep{hu23}, which means our previous finding can be reproduced.

Moreover, since the morphological stability of open clusters evolves over time, it is necessary to explore the relationship between the morphological stability of sample clusters on different projected planes and the age of the clusters. Figure~\ref{fig: ration_Age} shows the distribution between the morphological stability of sample clusters and their age, as well as its linear fitting. The age parameters of the sample clusters used in this work are from \citet{van23}. However, we do not find any linear correlation between these two parameters, which can also be verified by the \texttt{p-value} in Table.~\ref{table:Correlation}. The \texttt{p-value} of these relationships is more than~0.05. This means that the morphological stability of the sample clusters on different projection planes is independent of their age. In addition, the ratios of the age ranges of the sample clusters with morphological stability greater than 1 to the total age range of the entire sample clusters are 67.05\% for the X-Y plane, 71.59\% for X-Z plane, and 72.73\% for Y-Z plane, respectively. This means that our sample clusters do not have relatively stable projected morphologies at certain ages, especially for particularly young and old sample clusters. This may be because young clusters have not yet stabilized their morphology after birth, while older clusters have evolved over a long period of time and their morphology might have become unstable.

\subsection{Morphological stability of the 3D projected distribution of sample clusters vs. velocity dispersion}

\begin{figure*}
	\centering
	\includegraphics[angle=0,width=50mm]{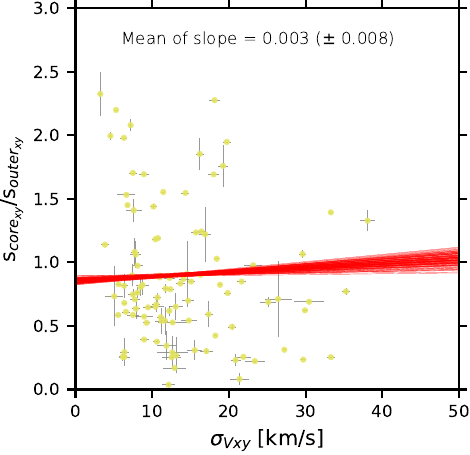}
	\includegraphics[angle=0,width=50mm]{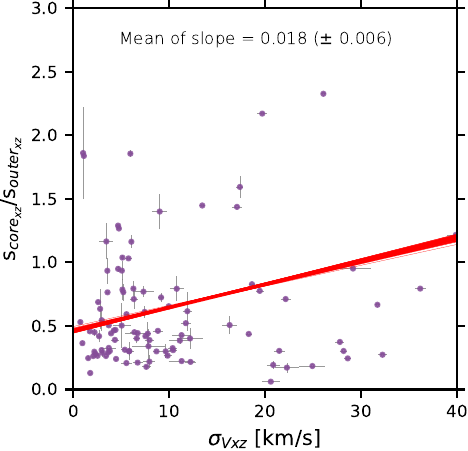}
	\includegraphics[angle=0,width=50mm]{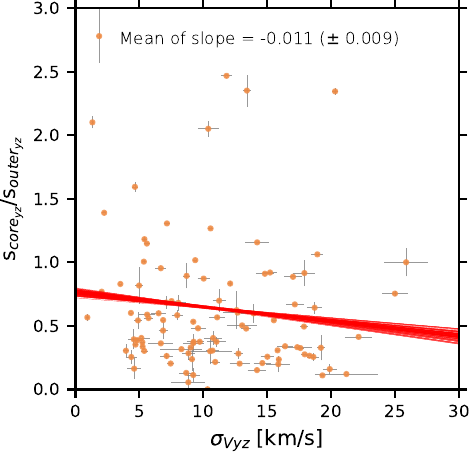}
	\caption{Morphological stabilities on the three planes within tidal radii vs. velocity dispersions on the three projections. Left: the morphological stability on the X-Y plane within tidal radii vs. $\sigma_{Vxy}$; Middle: the morphological stability on the X-Z plane within tidal radii vs. $\sigma_{Vxz}$; Right: the morphological stability on the Y-Z plane within tidal radii vs. $\sigma_{Vyz}$. Different colored scatters with gray error bars represent the sample clusters. The meaning of the values in the labels, symbols, and curves of the picture are the same as those of Fig.~\ref{fig: ration_N}. It should be noted that the number of samples in each panel is not equal to that of all samples. Because we removed those samples with the central core area ($S_{core_{xy}}$, or $S_{core_{xz}}$, or $S_{core_{yz}}$) being zero.}
	\label{fig: V_ratio}
\end{figure*}

The member stars of open clusters are in perpetual motion, so their morphological stability could potentially be associated with the velocity of their members, especially velocity dispersion. In this section, we investigate the correlation between the morphological stability of the 3D projected distributions of sample clusters and their corresponding projected velocity dispersion. However, the radial velocity data for the member stars of the sample clusters are not included in the dataset we initially downloaded. To address this, we matched the radial velocity data from the Gaia-DR3 database, using the TOPCAT software \citep{tayl05}. Finally, we utilized Astropy \citep{astr13, astr18} to calculate their kinematic velocities within the 3D projected distributions, and then obtained their velocity dispersion ($\sigma_{Vxy}$, $\sigma_{Vxz}$, and $\sigma_{Vyz}$). We note that due to the fact that not every member star of the sample clusters have radial velocity data, and that member stars with radial velocities may have large radial velocity errors. Therefore, we chose member stars with radial velocities and radial velocities greater than or equal to three times the radial velocity error to be used in the calculation of the dispersion of projected velocities of each cluster in three dimensions. We still calculated its error using the Monte Carlo method of sampling 100 times. These velocity dispersion values and their errors are summarized in Table~\ref{table:parameters_three} in the Appendix.

\begin{figure*}
	\centering
	\includegraphics[angle=0,width=50mm]{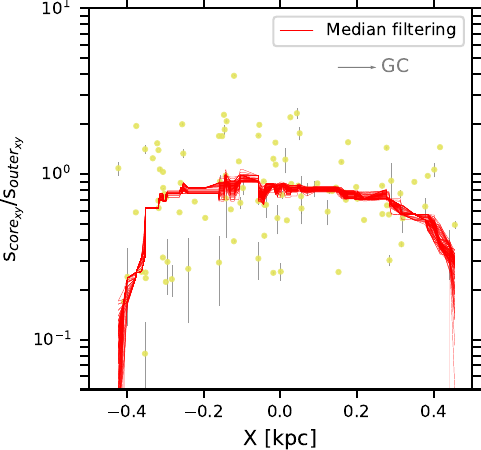}
	\includegraphics[angle=0,width=50mm]{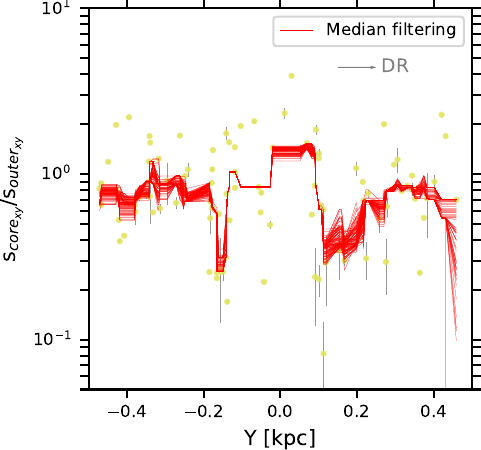}
	\includegraphics[angle=0,width=50mm]{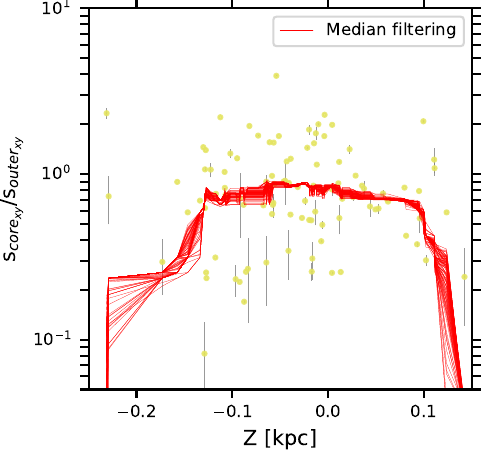}
	\caption{Distributions between the morphological stability of the projected distribution within tidal radii of sample clusters on the X-Y plane and the three axes (X axis, Y axis, and Z axis). `GC' and `DR' represent the Galactic center and the Galactic differential rotation, respectively. The colored scatters with gray error bars denote our sample clusters, with the red curve lines of each panel being the median filtering lines to the distribution of the panel. These red median filtering lines are obtained by sampling the parameters distributed in the figure 100 times using a Gaussian distribution within the range of error and filtering the median of the parameter distribution for each of these 100 sampling times. It should be noted that the number of samples in each panel is not equal to that of all samples. Because we removed those samples with the central core area ($S_{core_{xy}}$) being zero.}
	\label{fig: XYZ_ration_xy}
	
\end{figure*}

Figure~\ref{fig: V_ratio} presents the distributions comparing the morphological stabilities of sample clusters on the X-Y, X-Z, and Y-Z projections with their respective velocity dispersion on these planes. We performed 100 times of linear fit to these distributions, respectively. We find a linear relationship between the morphological stability on the X-Z plane and the velocity dispersion on the same plane only, with no significant linear relationships on other planes, which can be also verified by their correlation parameters in Table.~\ref{table:Correlation}. However, it is apparent that the number of samples with velocity dispersion greater than 20~\texttt{km/s} in the X-Z plane is small in our sample. Thus, we need to treat this linear relationship with caution. When not considering the influence of sample size, the results in the X-Y and Y-Z planes presented in Fig.~\ref{fig: V_ratio} contradicted our initial hypothesis. This may suggest that the size of the velocity dispersion within the sample clusters is not directly related to their morphological stability in the X-Y and Y-Z planes. Moreover, the morphological stability of clusters in the X-Z projection plane shows a linear positive correlation with velocity dispersion in the same plane, which may suggest that the more morphologically stable the sample cluster is in the X-Z projection plane, the greater its velocity dispersion. Therefore, we postulate that the morphological stability of the sample clusters may be related to the dynamical processes inside the clusters. We caution the readers that this conclusion may be a result of a limited sample size. Further research with a more extensive dataset is necessary to substantiate this finding.

\subsection{Morphological stability of the 3D projection distribution of sample clusters vs. positions}\label{Sect:position}

\begin{figure*}
	\centering
	\includegraphics[angle=0,width=50mm]{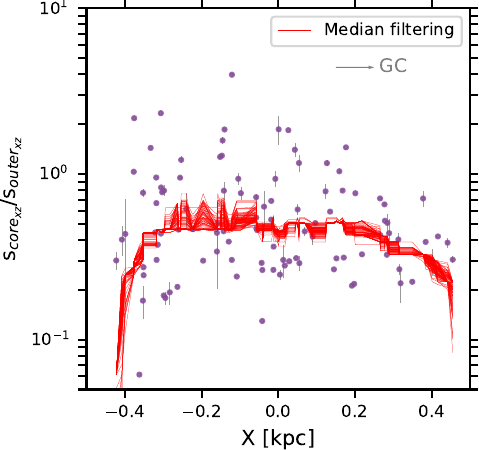}
	\includegraphics[angle=0,width=50mm]{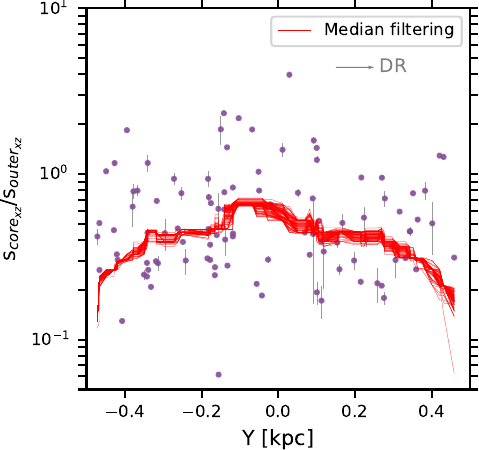}
	\includegraphics[angle=0,width=50mm]{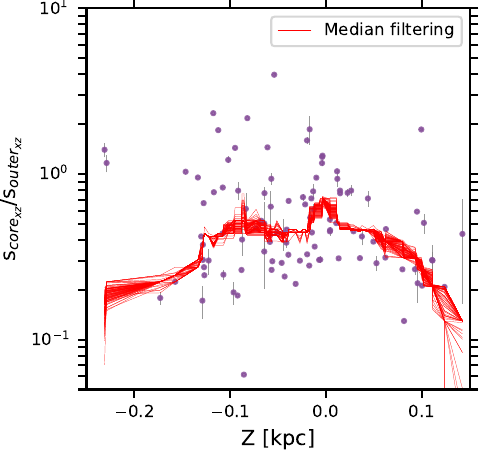}
	\caption{Distributions between the morphological stability of the projected distribution within tidal radii of sample clusters on the X-Z plane and the three axes (X axis, Y axis, and Z axis). `GC' and `DR' represent the Galactic center and the Galactic differential rotation, respectively. The colored scatters with gray error bars denote our sample clusters. The symbols and curves of the picture are the same as those of Fig.~\ref{fig: XYZ_ration_xy}. It should be noted that the number of samples in each panel is not equal to that of all samples. Because we removed those samples with the central core area ($S_{core_{xz}}$) being zero.}
	\label{fig: XYZ_ration_xz}
\end{figure*}

Sample clusters are embedded within the galactic disk and are inevitably subject to the gravitational potentials in the vicinity of their positions within the disk \citep{miho16, ange23}. In this study, our analysis is conducted within a heliocentric Cartesian coordinate system, with the Sun at its center. The positive X axis which extends from the position of the Sun's projection on the Galactic midplane toward the Galactic center is subject to a stronger potential than the negative X axis. This disparity could theoretically affect the morphological stability of the sample clusters as observed in the X-Y and X-Z projected planes. Regarding the Y-axis, its positive direction aligns with the rotational direction of the Galaxy. Assuming that the potential energy of the galactic disk is symmetrical about the X axis, it follows that the potential energy at the positive end of the Y axis should be symmetrical to that at the negative end, theoretically speaking. Under this assumption, the morphological stability of the sample clusters' projected distributions on both the X-Y and Y-Z planes should also reflect this symmetry. Likewise, the same assumption of potential symmetry can be extended to the Z axis as well.

\begin{figure*}
	\centering
	\includegraphics[angle=0,width=50mm]{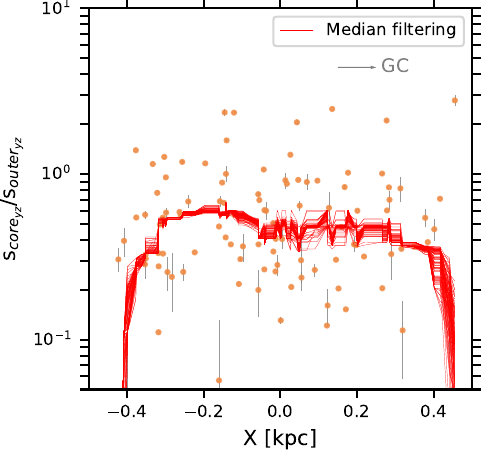}
	\includegraphics[angle=0,width=50mm]{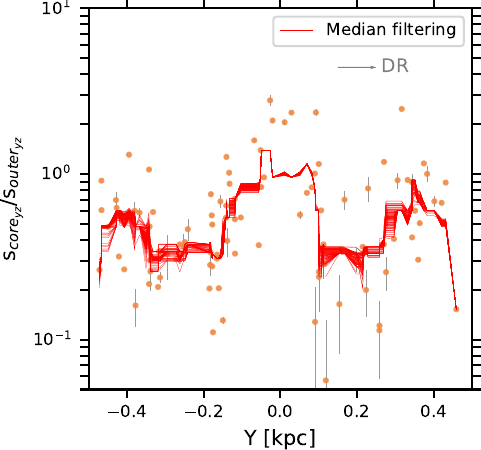}
	\includegraphics[angle=0,width=50mm]{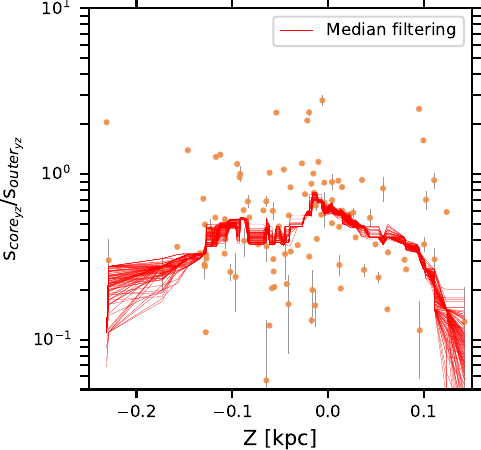}
	\caption{Distributions between the morphological stability of the projected distribution within tidal radii of sample clusters on the Y-Z plane and the three axes (X axis, Y axis, and Z axis). `GC' and `DR' represent the Galactic center and the Galactic differential rotation, respectively. The colored scatters with gray error bars denote our sample clusters. The symbols and curves of the picture are the same as those of Fig.~\ref{fig: XYZ_ration_xy}. It should be noted that the number of samples in each panel is not equal to that of all samples. Because we removed those samples with the central core area ($S_{core_{yz}}$) being zero.}
	\label{fig: XYZ_ration_yz}
\end{figure*}

To test our hypotheses, we also performed 100 times linear fit to the distributions between the morphological stability on the three projected planes and the three principal axes (X, Y, and Z), respectively. However, we do not find any significant linear correlation between them. Then to more clearly delineate nonlinear trends within these distributions, we applied the median filtering method (the \texttt{medfilt} function of \texttt{scipy.signal} \texttt{python} package). This technique yielded the median filtering line, depicted by the red curve in Figs.~\ref{fig: XYZ_ration_xy}, ~\ref{fig: XYZ_ration_xz}, and ~\ref{fig: XYZ_ration_yz} which is also utilized for subsequent distributions. We graphed the relationship between the morphological stabilities of the 3D projected distributions within tidal radii of the sample clusters and the three principal axes (X, Y, and Z), creating median filter lines for each distribution 100 times as displayed in Figs.~\ref{fig: XYZ_ration_xy}, \ref{fig: XYZ_ration_xz}, and \ref{fig: XYZ_ration_yz}. The left panel of Fig.~\ref{fig: XYZ_ration_xy} reveals that the morphological stability of the sample clusters on the X-Y projection does not significantly vary as we move along the X axis from negative to positive values. This suggests that the change in gravitational potential from the negative to the positive X axis, pointing toward the Galactic center, does not directly influence the morphological stability of the sample clusters on the X-Y projection plane. This finding is somewhat unexpected, as one would anticipate that the gravitational potential energy would increase with proximity to the Galactic center, thereby affecting the morphological stability. Consequently, clusters closer to the Galactic center would theoretically exhibit less morphological stability on the X-Y plane, compared to those situated further away. Of course, it is possible that the signal is not presented here because of the insufficient number of samples. In addition, there is another reason that the tidal structure of some clusters has not been totally and tidally disrupted, which may affect the direct connection of the morphological stability with the gravitational potential environment. Then this may result in the lack of the relationship between them. Thus, we still have to verify this conclusion in a large sample study.

Not coincidentally, the morphological stability of the sample clusters on the X-Y projection plane does not show a significant correlation with the Z axis, as depicted in the right panel of Fig.~\ref{fig: XYZ_ration_xy}. This outcome aligns with our expectations, given that the Z axis is orthogonal to the X-Y plane, and thus, the morphological stability within this plane should not be influenced by variations along the Z axis. Interestingly, the morphological stability of the sample clusters on the X-Y projection plane displays a discernible, regular fluctuation pattern along the Y axis, as seen in the middle panel of Fig.~\ref{fig: XYZ_ration_xy}. This fluctuation appears to be roughly symmetric around the point where the Y coordinate equals zero. This symmetry may suggest that the gravitational potentials on either side of the Y axis are indeed symmetrical and that the gravitational potential gradient along the Y axis exerts a nonuniform impact on the morphological stability of the clusters as observed in the X-Y projection. Of course, this fluctuation does not exhibit perfect symmetry, due to the fact that open clusters are clustered almost all around the Milky Way's spiral arms, while the latest studies \citep{hunt23, cava24, cant24} find that the spiral arms within 2~kpc, traced by young clusters, show a fragmented pattern. This may result in an irregular change in the gravitational potential along the Y axis. The reason for the result may be because the region close to the Sun is slightly further away from the local arm and without a more complex external environment, hence the clusters in this region are less perturbed and have a stronger morphological stability of their own. On the other hand, the region slightly farther away from the Sun is near the spiral arm, and the external force environment is complicated, so the influence on the clusters in this region is larger, and naturally their morphology stability is weaker.

Figure~\ref{fig: XYZ_ration_xz} illustrates the relationship between the morphological stability of the projected distributions of the sample clusters on the X-Z plane and the three axes (X, Y, and Z). Given that the X-Z projection plane is orthogonal to the Y axis, the morphological stability on this plane should be independent of the Y axis. This expectation is supported by the median panel of Fig.~\ref{fig: XYZ_ration_xz}, where no correlation between the morphological stability on the X-Z plane and the Y axis is observed. Furthermore, akin to the stability on the X-Y plane, the morphological stability on the X-Z plane also appears to be unrelated to the X axis, as shown in the left panel of Fig.~\ref{fig: XYZ_ration_xz}. This result might be attributed to sample effects. Alternatively, it could be that 
the gravitational potential along the X axis, particularly from the negative to the positive direction toward the Galactic center, may not be strong enough to affect the morphological stability of the sample clusters on the X-Z plane, while it likely still affects the parts out of their gravitationally bound region. 

Nevertheless, a modest fluctuation is observed between the morphological stability on the X-Z plane and the Z axis, as depicted in the right panel of Fig.~\ref{fig: XYZ_ration_xz}. This suggests that along the Z axis, there is a variation in gravitational potential, which in turn induces a slight fluctuation in the morphological stability of the sample clusters on the X-Z projection. It is important to note that this observation could also be a consequence of sample effects. We call for further validation through a large scale study.

Moreover, we examined the relationship between the morphological stability of the projected distributions within tidal radii of sample clusters on the Y-Z plane and the three axes (X, Y, and Z), as depicted in Fig.~\ref{fig: XYZ_ration_yz}. The left panel of Fig.~\ref{fig: XYZ_ration_yz} confirms that the morphological stability on the Y-Z projection plane is unrelated to the X-axis. This is expected, as the X axis is perpendicular to the Y-Z plane, and thus, variations in gravitational potential along the X axis do not influence the morphological stability of the sample clusters within this plane. Interestingly, the morphological stability on the Y-Z projection plane exhibits a regular fluctuation along the Y axis, which is approximately symmetric around the Y = 0 position, as shown in the middle panel of Fig.~\ref{fig: XYZ_ration_yz}. This pattern may suggest that the gravitational potentials on either side of the Y axis produce regular changes, leading to the observed oscillations in morphological stability. However, given the small sample size in this study, this finding requires further verification with a larger dataset. Furthermore, a subtle fluctuation is observed between the morphological stability on the Y-Z projection plane and the Z axis, as shown in the right panel of Fig.~\ref{fig: XYZ_ration_yz}. There seems to be a tendency for sample clusters close to the galactic plane to be more morphologically stable on the Y-Z plane than that of sample clusters farther away from the galactic disk plane. This may be due to the strong vertical tidal field of the disk \citep{mart17}. Since our sample clusters have a limited distribution along the Z axis, this fluctuation could potentially be a result of sample effects. Thus, we need to treat this small fluctuation cautiously.

\subsection{Morphological stability of the 3D projected distribution of sample clusters vs. external environment}

\begin{figure*}
	\centering
	\includegraphics[angle=0,width=50mm]{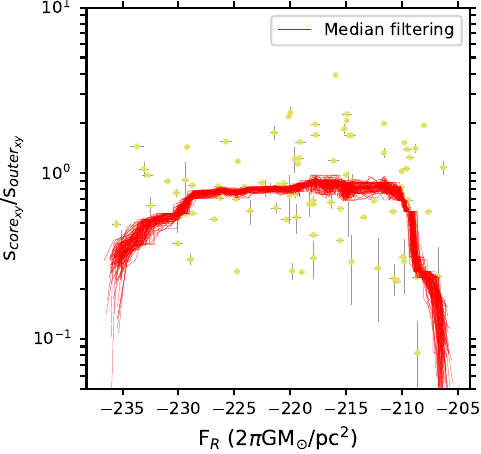}
	\includegraphics[angle=0,width=50mm]{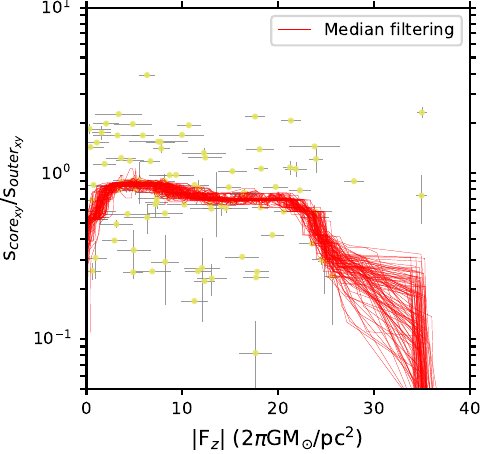}
	\includegraphics[angle=0,width=50mm]{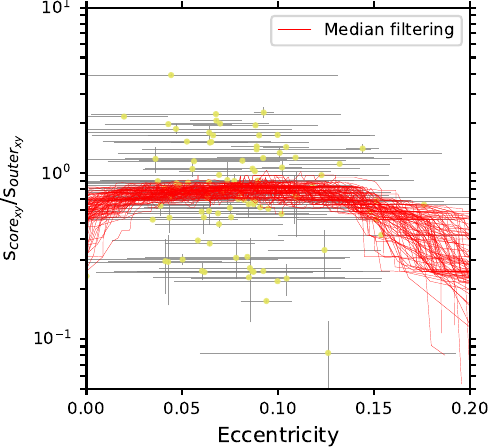}
	\caption{Morphological stability on the X-Y plane within tidal radii vs. three parameters of external environment. Left: the morphological stability on the X-Y plane within tidal radii vs. $F_{R}$; Middle: the morphological stability on the X-Y plane within tidal radii vs. the absolute value of $F_{z}$; Right: the morphological stability on the X-Y plane within tidal radii vs. Eccentricity. The colored scatters with gray error bars denote our sample clusters. The symbols and curves of the picture and the number of samples are the same as those of Fig.~\ref{fig: XYZ_ration_xy}.}
	\label{fig: Forces_ration_xy}
\end{figure*}

\begin{figure*}
	\centering
	\includegraphics[angle=0,width=50mm]{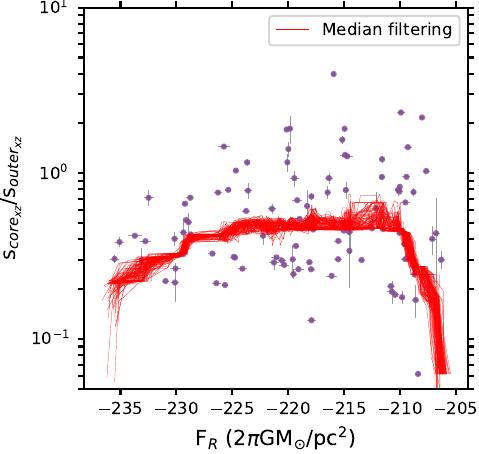}
	\includegraphics[angle=0,width=50mm]{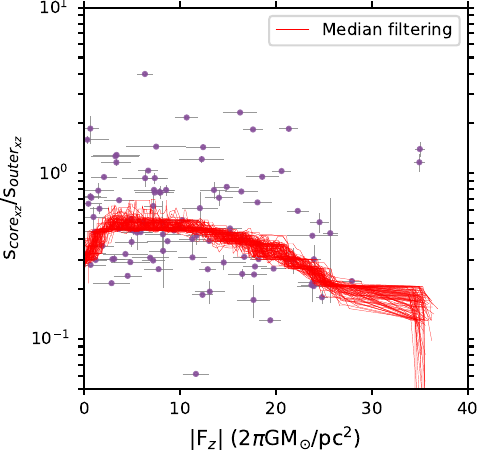}
	\includegraphics[angle=0,width=50mm]{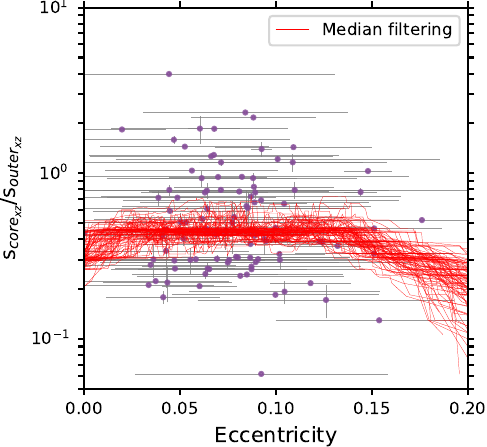}
	\caption{Morphological stability on the X-Z plane within tidal radii vs. three parameters of external environment. Left: the morphological stability on the X-Z plane within tidal radii vs. $F_{R}$; Middle: the morphological stability on the X-Z plane within tidal radii vs. the absolute value of $F_{z}$; Right: the morphological stability on the X-Z plane within tidal radii vs. Eccentricity. The colored scatters with gray error bars denote our sample clusters. The symbols and curves of the picture and the number of samples are the same as those of Fig.~\ref{fig: XYZ_ration_xy}.}
	\label{fig: Forces_ration_xz}
\end{figure*}

\begin{figure*}
	\centering
	\includegraphics[angle=0,width=50mm]{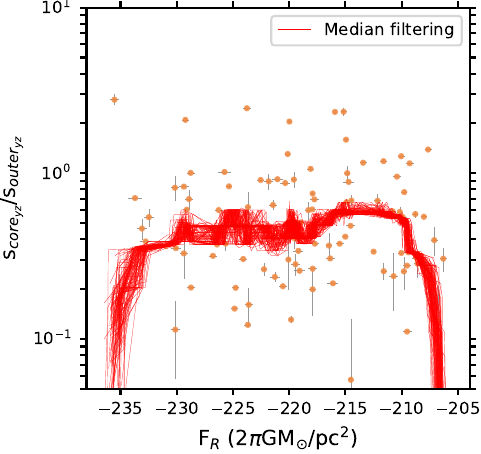}
	\includegraphics[angle=0,width=50mm]{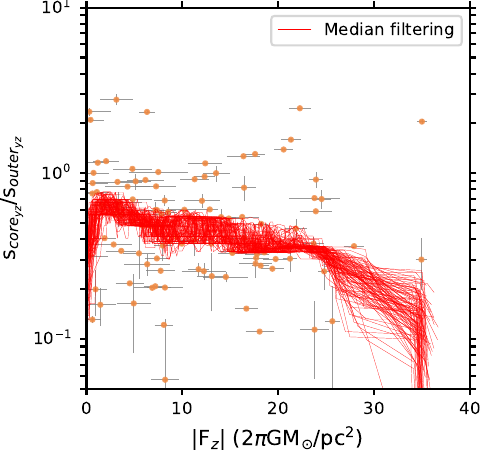}
	\includegraphics[angle=0,width=50mm]{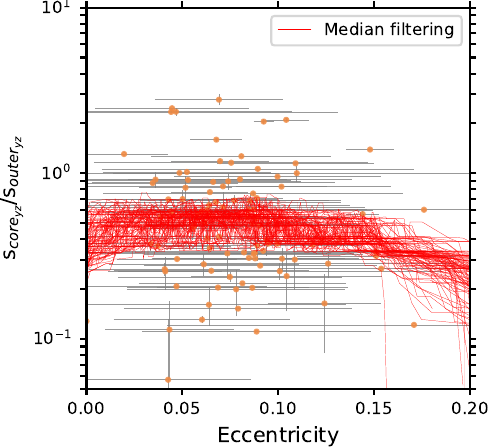}
	\caption{Morphological stability on the Y-Z plane within tidal radii vs. three parameters of external environment. Left: the morphological stability on the Y-Z plane within tidal radii vs. $F_{R}$; Middle: the morphological stability on the Y-Z plane within tidal radii vs. the absolute value of $F_{z}$; Right: the morphological stability on the Y-Z plane within tidal radii vs. Eccentricity. The colored scatters with gray error bars denote our sample clusters. The symbols and curves of the picture and the number of samples are the same as those of Fig.~\ref{fig: XYZ_ration_xy}.}
	\label{fig: Forces_ration_yz}
\end{figure*}

In Sect.~\ref{Sect:position}, we observe that, notwithstanding the influence of sample effects, the morphological stability of the projected distributions within tidal radii of the sample clusters across three orthogonal planes might be correlated with their spatial positions along the Y axis. Given that the external environment fluctuates across different areas, this section is dedicated to investigating a potential direct connection between the morphological stability of the 3D projected distributions of the sample clusters and the external environment \citep{mart17}. To achieve this, we primarily utilized the \texttt{galpy.potential} module, specifically the \texttt{MWPotential2014} \citep{miya75, nava97}, from the Python package \texttt{galpy} \citep{bovy15}. This module facilitates the extraction of various parameters indicative of the influence of the external environment. Under this framework, we are able to compute the radial forces (F$_{R}$) and vertical forces (F$_{z}$) in cylindrical coordinates\footnote{These are cylindrical coordinates centered on the Galactic center, adhering to the left handed frame convention as per \texttt{galpy}.}, as well as the orbit eccentricities of the sample clusters, as detailed in Table~\ref{table:parameters_two} in the Appendix. This is accomplished by inputting their 2D celestial coordinates, proper motions, and radial velocities. For an in-depth understanding of these parameters, we direct the readers to the web\footnote{\url{https://docs.galpy.org/en/v1.9.2/}}, which provides a comprehensive overview of these parameters.

Figures~\ref{fig: Forces_ration_xy}, \ref{fig: Forces_ration_xz}, and \ref{fig: Forces_ration_yz} show the distributions relating the morphological stabilities of the projected distributions (on the X-Y, X-Z, and Y-Z planes, respectively) within tidal radii of sample clusters to external parameters, including radial force ($F_{R}$), vertical force ($F_{z}$), and orbital eccentricity. We drew median filter lines in all these distributions. Given that $F_{R}$ is parallel to the X-Y and X-Z planes, it is theoretically expected to predominantly influence the morphological stabilities of the sample clusters on these two planes. The left panels of Figs.~\ref{fig: Forces_ration_xy} and \ref{fig: Forces_ration_xz} show that the morphological stability of the sample clusters in the X-Y and X-Z planes increases slightly to stabilization and then decreases sharply as the $F_{R}$ value decreases. These changes are consistent with our theoretical expectations. However, the sharp decrease is most likely due to the boundary effect of the samples. This trend may show that a reduced influence of $F_{R}$ on the X-Y and X-Z planes correlates with greater morphological stability for the sample clusters. However, it is important to acknowledge that these slight upward trends are likely influenced by sample effects, and thus, the intriguing findings necessitate further validation. Additionally, the left panel of Fig.~\ref{fig: Forces_ration_yz} demonstrates that there is no relationship between the morphological stability on the Y-Z projection plane and $F_{R}$. This result is logically sound, as $F_{R}$ is perpendicular to the Y-Z plane, and consequently, does not affect the morphological stability of the sample clusters within this plane.

Furthermore, we observe downward trends between the vertical force ($F_{z}$) and the morphological stabilities of sample clusters on the two projection planes (X-Z and Y-Z planes), as depicted in the median panels of Figs.~\ref{fig: Forces_ration_xz}, and~\ref{fig: Forces_ration_yz}. These trends imply that the morphological stability of the sample clusters tends to decrease with increasing $F_{z}$. This is readily explainable. Given that $F_{z}$ is parallel to the X-Z and Y-Z planes, there is a direct impact on the morphological stability of the sample clusters projected onto these planes. Moreover, we find from the median panel of Fig.~\ref{fig: Forces_ration_xy} a slightly downward trend between the vertical force ($F_{z}$) and the morphological stability of sample clusters on the X-Y projection plane. But this is most likely due to sample effects. Meanwhile, this trend is somewhat counterintuitive, as one would theoretically expect the morphological stability on the X-Y plane to remain unaffected by $F_{z}$ that is perpendicular to the X-Y plane.

Subsequently, we investigate the potential relationship between the morphological stability of the sample clusters on the three projection planes (X-Y, X-Z, and Y-Z) and their orbital eccentricity. Upon examining the right panels of Figs.~\ref{fig: Forces_ration_xy},~\ref{fig: Forces_ration_xz}, and~\ref{fig: Forces_ration_yz}, we do not discern any evident correlation between the morphological stability of the sample clusters and their orbital eccentricity. This suggests that the variations in orbital potentials associated with different eccentricities may not directly influence the morphological stability of the 3D projected distributions of the sample clusters.

\section{Summary}

In this study, we employed the rose diagram method to quantify the 3D projected morphology of 105 nearby open clusters selected from the literature. Additionally, we defined the morphological stability of these clusters' 3D projected distributions within tidal radii for the first time and investigated some factors that may influence this stability in different projected planes. Our key findings are as follows:

1. There is a positive correlation between the morphological stability of the sample clusters on different projection planes and the number of their member stars within tidal radii. This may indicate that the more member stars within the tidal radius inside a cluster, the stronger its own gravitational binding ability, resulting in a very stable morphological stability.

2. The morphological stability of the sample clusters on the X-Y and Y-Z projection planes within tidal radii does not appear to be directly related to their corresponding velocity dispersion. However, there is a positive linear relationship between the morphological stability in the X-Z plane and the velocity dispersion in the same plane, suggesting that the morphological stability in the X-Z plane may have a link to the internal dynamics of clusters.

3. The morphological stability of the 3D projected distributions within tidal radii of the sample clusters shows a degree of correlation with their spatial locations along the Y axis. This implies that changes in the gravitational potentials at the clusters' locations along the Y axis may directly affect their morphological stabilities in the 3D projected planes within tidal radii.

4. Specific external forces within the clusters' environments may have a direct impact on the morphological stability of their 3D projected distributions within tidal radii.

Our research quantifies of the morphological stability of open clusters on 3D projected planes and offers new insights into their 3D projected morphologies.

\begin{acknowledgements}
	We thank the reviewer for a very detailed comment report that made this article significantly better. This work is supported by the National Natural Science Foundation of China (NSFC) under grant 12303037, the Fundamental Research Funds of China West Normal University (CWNU, No.493065), the National Key R\&D Program of China (Nos. 2021YFA1600401 and 2021YFA1600400), the National Natural Science Foundation of China (NSFC) under grant 12173028, the Chinese Space Station Telescope project: CMS-CSST-2021-A10, the Sichuan Youth Science and Technology Innovation Research Team (Grant No. 21CXTD0038), and the Innovation Team Funds of China West Normal (No. KCXTD2022-6). Qingshun Hu would like to acknowledge the financial support provided by the China Scholarship Council program (Grant No. 202308510136). S. Q. acknowledges the financial support provided by the China Scholarship Council program (Grant No. 202304910547). This study has made use of the Gaia DR3, operated by the European Space Agency (ESA) space mission (Gaia). The Gaia archive website is \url{https://archives.esac.esa.int/gaia/}.

	Software: Astropy \citep{astr13,astr18}, Scipy \citep{gomm24}, Numpy \citep{harr20}, TOPCAT \citep{tayl05}, Galpy \citep{bovy15}.
	
\end{acknowledgements}



\begin{appendix}

\section{Robustness of the symmetric signals between the morphological stability on the X-Y and Y-Z planes and the Y-axis}

\begin{figure}[ht]
	\centering
	\includegraphics[width=45mm]{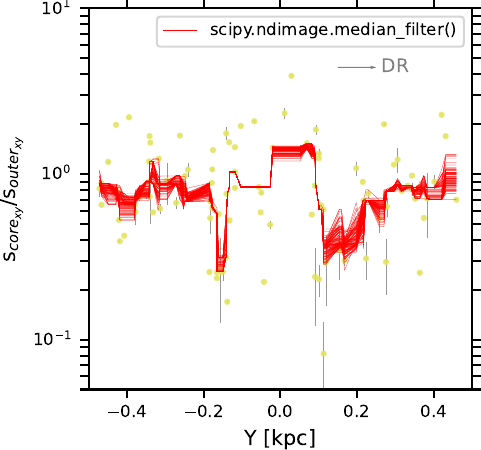}
	\includegraphics[width=45mm]{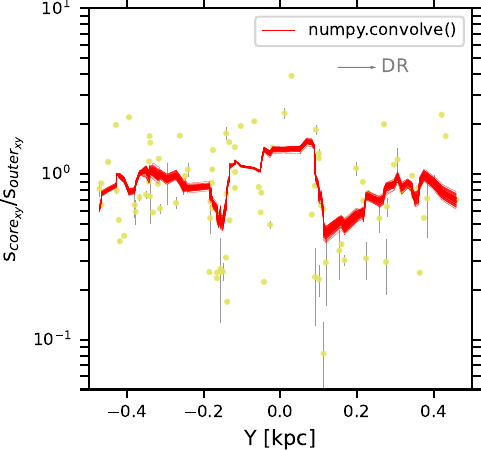}
	\includegraphics[width=45mm]{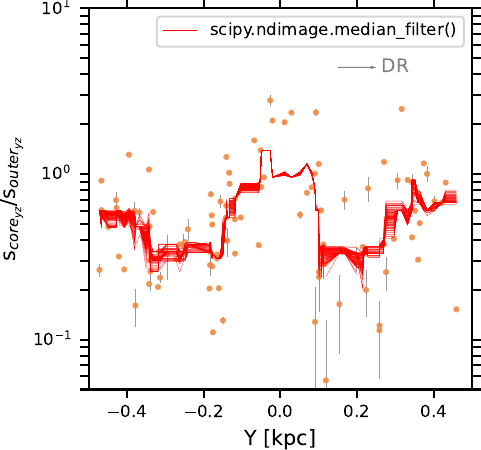}
	\includegraphics[width=45mm]{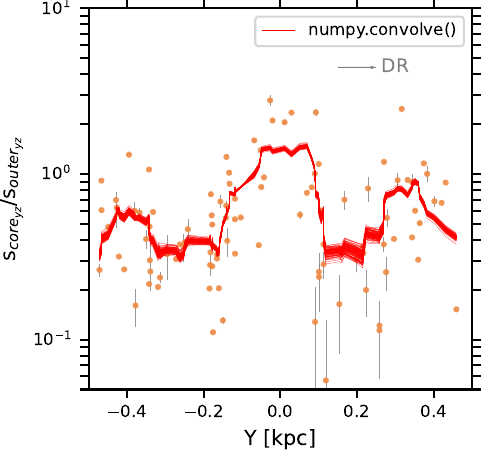}
	\caption{The symmetries between the morphological stability of our sample clusters on the X-Y and Y-Z projected planes and the Y-axis, obtained with the median and mean filtering methods.}
	\label{Median_mean_filtering}
\end{figure}

We adopted the two methods to verify the symmetric signals between the morphological stability of our sample clusters on the X-Y and Y-Z projected planes and the Y-axis. One of them is a median filtering method (the \texttt{median\_filter} function of \texttt{scipy.ndimage} \texttt{python} package), with the other one being a mean filtering (the \texttt{convolve} function of \texttt{numpy} \texttt{python} package). Figure~\ref{Median_mean_filtering} shows the symmetric trends presented in the distribution between the morphological stability on the X-Y and Y-Z projected planes and the Y-axis, using the median and mean filtering, respectively. This indicates that the symmetries in our work are stable.

\begin{sidewaystable}[ht]
	\caption{The morphological stability and position parameters of the sample clusters}\label{table:parameters_one}
	\centering
	\small
	    \begin{tabular}{lccccccccc}
		
		\hline\noalign{\smallskip}
		\hline\noalign{\smallskip}
		
		(1) & (2) & (3) & (4) & (5) & (6) & (7) & (8) & (9) & (10) \\
		Name &  X & Y  & Z  &  $S_{core_{xy}}$/$S_{outer_{xy}}$ &  $\sigma_{(S_{core_{xy}}/S_{outer_{xy}})}$  & $S_{core_{xz}}$/$S_{outer_{xz}}$ &  $\sigma_{(S_{core_{xz}}/S_{outer_{xz}})}$ & $S_{core_{yz}}$/$S_{outer_{yz}}$ &  $\sigma_{(S_{core_{yz}}/S_{outer_{yz}})}$ \\
		
		\hline\noalign{\smallskip}
		
		-	&  pc & pc &  pc & - &  - & -  & -  & - &- \\
		
		\hline\noalign{\smallskip}
		
		NGC 2547 & -37.13 & -379.57 & -57.81 & 0.65 & 0.11 & 0.63 & 0.16 & 0.6 & 0.07 \\
		NGC 3532 & 159.12 & -448.68 & 11.31 & 1.18 & 0.01 & 1.04 & 0.02 & 0.48 & 0.0 \\
		Platais 10 & 151.6 & -184.79 & 12.94 & 0.26 & 0.0 & 0.31 & 0.0 & 0.2 & 0.0 \\
		NGC 1333 & -253.05 & 100.58 & -102.17 & 1.33 & 0.08 & 1.22 & 0.06 & 0.26 & 0.03 \\
		Turner 5 & -41.79 & -406.99 & 81.13 & 0.42 & 0.0 & 0.13 & 0.0 & 0.27 & 0.0 \\
		Blanco 1 & 43.14 & 11.14 & -231.14 & 2.33 & 0.18 & 1.4 & 0.14 & 2.05 & 0.06 \\
		UPK 305 & -318.06 & 216.6 & -133.82 & 0.69 & 0.0 & 0.95 & 0.0 & 0.33 & 0.0 \\
		
		... & ...& ... & ...& ... & ... & ...& ... & ... & ... \\
		
		\hline\noalign{\smallskip}
	\end{tabular}
	\tablefoot{Column~(1) lists the name of the sample clusters. Columns~(2), (3), and (4) correspond to the X, Y, and Z coordinates of the clusters, respectively. Columns~(5), (7), and (9) represent the morphological stabilities on the X-Y, X-Z, and Y-Z projection planes within tidal radii, respectively, with Columns~(6), (8), and (10) represent the error of the morphological stabilities on the X-Y, X-Z, and Y-Z projection planes, respectively. This is a machine-readable table, and the complete dataset is available for consultation through the CDS.}
	\flushleft
	
\end{sidewaystable}

\begin{sidewaystable}[ht]
	\caption{The external forces and fundamental parameters of the sample clusters}\label{table:parameters_two}
	\centering
	\small

	\begin{tabular}{lcccccccccc}
		
		\hline\noalign{\smallskip}
		\hline\noalign{\smallskip}
		
		(1) & (2) & (3) & (4) & (5) & (6) & (7) & (8) & (9) & (10) & (11)\\
		Name &  N$_{rt}$ &  $\sigma_{N_{rt}}$  & R$_{t}$  & $\sigma_{R_{t}}$ & $F_{R}$ & $\sigma_{F_{R}}$ & $F_{z}$ & $\sigma_{F_{z}}$ & $E$ & $\sigma_{E}$ \\
		
		\hline\noalign{\smallskip}
		
		-	&  - & - &  pc & pc &  2$\pi$G$M_{\odot}$/$pc^{2}$ & 2$\pi$G$M_{\odot}$/$pc^{2}$ & 2$\pi$G$M_{\odot}$/$pc^{2}$ & 2$\pi$G$M_{\odot}$/$pc^{2}$ &- & - \\
		
		\hline\noalign{\smallskip}
		
		NGC 2547 & 617.0 & 28.0 & 19.33 & 3.26 & -218.29 & 0.29 & 7.18 & 0.92 & 0.08 & 0.06 \\
		NGC 3532 & 3096.0 & 20.0 & 24.88 & 1.42 & -224.68 & 0.23 & -6.66 & 1.05 & 0.06 & 0.05 \\
		Platais 10 & 98.0 & 0.0 & 200.0 & 0.0 & -224.76 & 0.3 & -6.86 & 1.94 & 0.09 & 0.05 \\
		NGC 1333 & 514.0 & 5.0 & 38.1 & 1.15 & -211.58 & 0.26 & 12.24 & 2.32 & 0.1 & 0.08 \\
		Turner 5 & 103.0 & 0.0 & 92.83 & 66.14 & -217.9 & 0.38 & -19.39 & 1.16 & 0.15 & 0.05 \\
		Blanco 1 & 640.0 & 8.0 & 13.42 & 0.8 & -219.97 & 0.19 & 34.98 & 0.52 & 0.09 & 0.01 \\
		UPK 305 & 150.0 & 0.0 & 197.55 & 14.53 & -209.4 & 0.36 & 18.54 & 1.7 & 0.08 & 0.09 \\
		
		... & ...& ... & ...& ... & ... & ...& ... & ... & ... & ...\\
		
		\hline\noalign{\smallskip}
	\end{tabular}
	\flushleft
	\tablefoot{Column~(1) represents the name of sample clusters. Columns~(2) and (3) lists the count of member stars per cluster within the tidal radius and its error, respectively. Columns~(4) and (5) refer to the tidal radii of the sample clusters and their errors, respectively. Columns~(6) and (7) represent the radial force and its uncertainty, while columns~(8) and (9) denote the vertical force and its error. Columns~(10) and (11) are the orbital eccentricity and its error, respectively. This table is presented in a machine-readable format, with the complete dataset accessible at the CDS.}
	
\end{sidewaystable}

\begin{sidewaystable}[ht]
	\caption{The velocity dispersion and other parameters of the sample clusters}\label{table:parameters_three}
	\centering
	\small
	
	\begin{tabular}{lccccccccccccc}
		
		\hline\noalign{\smallskip}
		\hline\noalign{\smallskip}
		
		(1) & (2) & (3) & (4) & (5) & (6) & (7) & (8) & (9) & (10) & (11) & (12) & (13) & (14)\\
		Name &  $\sigma_{Vxy}$ & $\sigma_{\sigma_{Vxy}}$ & $\sigma_{Vxz}$ & $\sigma_{\sigma_{Vxz}}$ & $\sigma_{Vyz}$ & $\sigma_{\sigma_{Vyz}}$ & N$_{Rv}$ & $r_{xy}$ & $\sigma_{r_{xy}}$ & $r_{xz}$ & $\sigma_{r_{xz}}$ & $r_{yz}$ & $\sigma_{r_{yz}}$ \\
		
		\hline\noalign{\smallskip}
		
		-	&  km/s & km/s &  km/s & km/s & km/s & km/s & - &  - & - & - & - &- & - \\
		
		\hline\noalign{\smallskip}
		
		NGC 2547 & 13.07 & 0.93 & 2.79 & 0.13 & 13.92 & 0.95 & 95 & 0.58 & 0.04 & 0.56 & 0.07 & 0.55 & 0.03 \\
		NGC 3532 & 10.41 & 0.41 & 5.11 & 0.11 & 13.35 & 0.41 & 209 & 0.69 & 0.0 & 0.81 & 0.01 & 0.54 & 0.0 \\
		Platais 10 & 6.37 & 0.75 & 10.4 & 0.45 & 12.84 & 0.78 & 23 & 0.4 & 0.0 & 0.35 & 0.0 & 0.34 & 0.0 \\
		NGC 1333 & 38.07 & 0.92 & 39.88 & 0.91 & 18.61 & 0.47 & 72 & 0.78 & 0.02 & 0.73 & 0.01 & 0.41 & 0.02 \\
		Turner 5 & 18.23 & 0.23 & 1.75 & 0.03 & 18.3 & 0.24 & 15 & 0.43 & 0.0 & 0.28 & 0.0 & 0.39 & 0.0 \\
		Blanco 1 & 3.23 & 0.19 & 9.0 & 0.81 & 10.38 & 0.81 & 68 & 0.94 & 0.02 & 0.85 & 0.04 & 0.87 & 0.01 \\
		UPK 305 & 30.44 & 1.93 & 29.16 & 1.83 & 17.34 & 1.16 & 9 & 0.62 & 0.0 & 0.69 & 0.0 & 0.42 & 0.0 \\
		
		... & ...& ... & ...& ... & ... & ...& ... & ... & ... & ... & ... & ... & ... \\
		
		\hline\noalign{\smallskip}
	\end{tabular}
	\flushleft
	\tablefoot{Column~(1) represents the name of sample clusters. Columns~(2), (4), and (6) are the velocity dispersion of sample clusters in the X-Y, X-Z, and Y-Z projected planes within tidal radii, respectively, with Columns~(3), (5), and (7) referring to the error of the velocity dispersion of sample clusters in the X-Y, X-Z, and Y-Z projected planes, respectively. Column~(8) lists the number of members per cluster with radial velocities that are used to calculate the velocity dispersion of the clusters. Additionally, columns~(9), (11), and (13) denote the radii of the central core regions of the clusters on the X-Y, X-Z, and Y-Z projection planes within tidal radii, respectively, with their errors being listed in columns~(10), (12), and (14). This table is presented in a machine-readable format, with the complete dataset accessible at the CDS.}
	
\end{sidewaystable}

\end{appendix}


\end{document}